# On the Performance Evaluation and Analysis of the Hybridised Bittorrent Protocol with Partial Mobility Characteristics

George C. Violaris[1], Constandinos X. Mavromoustakis[2]

*Computer Science Department, University of Nicosia*
*46 Makedonitissas Avenue, 1700 Nicosia, Cyprus*

[1]`violaris.g@student.unic.ac.cy`
[2]`mavromoustakis.c@unic.ac.cy`

*Abstract*— Engaging mobility with file sharing is considered very promising in today's run Anywhere, Anytime, Anything (3As) environments. The Bittorrent file sharing protocol can be rarely combined with the mobility scenario framework since resources are not available due to the dynamically changing topology network. As a result, mobility in P2P-oriented file sharing platforms, degrades the end-to-end efficiency and the system's performance. This work proposes a new hybridized model, which takes into account the mobility characteristics of the combined Bittorrent protocol in a centralized manner enabling partial mobility characteristics, where the clients of the network use a distinct technique to differentiate between mobile and static nodes. Many parameters were taken into consideration like the round trip delays, the diffusion process, and the seeding techniques, targeting the maximization of the average throughput in the clustered swarms containing mobile peers. Partial mobility characteristics are set in a peer-tracker and peer-peer communication enhancement schema with partial mobility, allowing an optimistic approach to attain high availability and throughput response as simulation results show.

## I. INTRODUCTION

In recent years, the Bittorrent protocol has become an increasingly successful method for delivering end-to-end data, with reliability and efficiency. The tit-for-tat techniques [1] which are built in the protocol require peers to seed back the content they have received. Many researches have been inspired in order to improve Bittorrent's performance [2][3][10] and [12]. Different scenarios and algorithms have been implemented and thoroughly tested [2][6][7][8][9], seeking ways to maximize the end-to-end performance using P2P techniques and approaches.

As in other P2P file-sharing schemes, performance depends mainly on the robustness of each node. Robustness depends on the temporal characteristics as well as on the spatial characteristics like whether the nodes are dynamically moving, etc. However there are certain features that need to be taken into consideration in order to enable higher performance onto a node-to-node sharing scenario. These, do not only rely on the behaviour of the connection between nodes, but on the techniques used to ensure quality of service through the protocol itself. In its current state, the protocol relies on the following ways, described by [2], to maintain the connectivity issue as follows:

- *Network size:* The number of peers in a Bittorrent network is important to determine metrics such as the request arrival rate, peer departure rate and the upload/download ratio in the bandwidth of each peer.

- *Efficient distribution:* Peers exchange pieces of a file, by a method called swarming [3]. In order to maintain efficiency, it is important to devise ways so peers do not get the same or very popular pieces. This is the reason the rarest-first policy [1], exists in the Bittorrent protocol; to maximise the potential of efficient distribution among peers.

- *Leech avoidance:* When a peer downloads without retaliation of the content they receive, the peer is called a leech. When there is a high ratio of these free-riding clients, the results are catastrophic for other peers. Therefore, mechanisms have been built to prevent this from happening *(one such example is the tit-for-tat algorithm, giving means to ensure fair transfers of data)*.

Enabling these devices with mobility characteristics and utilizing them with the Bittorrent protocol, many restrictions arise. Peers are prone to failures and aggravate the end-to-end performance, whereas short connections times or sudden disconnections *(with chained unpredictable disconnections due to range and battery failures)* reduce the overall resource availability of the MP2P system. Moreover mobile peers are subject to limited bandwidths, both in the download and upload activities. Additionally, the protocol specifications make use of the tit-for-tat policy [1], which essentially means equivalent retaliation of pieces amongst peers. Since mobile peers do not contribute dramatically to other peers due to their limitation in bandwidth, other peers will perceive them as leechers, and therefore they will avoid providing content to them. When speaking of mobile peers, the most common understanding of the term is about those devices which use a network wirelessly, i.e., cellular phones, however mobile phones which are connected through Wi-Fi are considered to have high bandwidth and therefore not regarded as mobile per se. Devices making use of GPRS or 3G standards, are the targeted technologies of this research.

The present work, proposes a new hybrid policy for the Bittorrent protocol using P2P strategies enabling nodes with partial mobility characteristics, where the clients of a network use a distinctive technique to differentiate between mobile and statically located nodes. The model has been devised in order to enable the seeding peers that will split the uploading portion of their bandwidths towards a higher number of mobile peers, in order to enable enhance network mobility. The scheme therefore can be hosted in larger scale Bittorrent clusters. The block-to-block and round trip delays are taken into consideration, enabling peer selection and seeding strategies to take place, targeting the maximisation of the average throughput in clusters containing mobile peers. The proposed scheme utilizes systems resources and comprises of a new model for disseminating information in a P2P system. The proposed scheme, hosts these partial mobility characteristics in a peer-to-tracker and peer-to-peer communication enhancement scheme, allowing an optimized approach to be applied for high resource availability in P2P networks with partial mobility characteristics.

The rest of our paper is organised as follows: Section 2 reviews previous work done on the Bittorrent protocol and similar static and non-static P2P methodologies. Section 3 provides information about the potential of mobility in Bittorrent, analysing the current problems mobile P2P transactions face and proceeds to explain the proposed hybridised model for dynamically changing topology systems, allowing mobility to peers. Section 4 discusses the simulation results and presents a performance analysis of the hybridised model, providing also discussion on seeding techniques and peer selection strategies. Finally, Section 5 concludes with a summary of the findings from the simulation study and discusses the future research directions the current research will take.

## II. RELATED WORK

Bittorrent performance is not only dependant to the protocol's peer selection algorithms and the tit-for-tat techniques. Certain simulation experimental studies show that along with optimised algorithms for content distribution, some minor alternations in the protocol's policies could significantly improve long term performance. Since the tit-for-tat policy of Bittorrent only takes place for a single file transfer at a specific moment in time [5], the sharing of old content is not rewarded and/or credited. Therefore incentives that elongate a content's lifetime are needed as files of high resource demands may become unavailable.

An analytical study in [2] has shown through a fluid model of the Bittorrent protocol that the average download time does not depend on the node arrival rate. Also, the study shows that there is a high chance that a peer will hold a specific block which other peers may be in need of. This concept allows for mobile clients to be 'optimistic' on having content delivered to them; however, some limitations which apply in the Bittorrent architecture do not enable these kinds of peers to use the full potential of their bandwidths.

It can be observed that Bittorrent uses a fixed default number, $u = 5$ reported in [4][6], of upload connections at any given time. The study reveals two significant problems. Firstly, the availability of full blocks to the network is delayed or postponed due to the high number of concurrent uploads occurring. Due to this, latency is significantly increased. Secondly, the seeding peer may be uploading to the downloading peer faster than the latter can receive blocks. This happens as the peer's bandwidth is congested on the downloading side, thus increasing the number of lost packets, leading to high redundancy in the network and unneeded repetition.

The simulation in [7] shows how Bittorrent works in general, while giving emphasis in super-seeding. Furthermore, the study shows how simulations can produce statistics for large scale experimentation that would otherwise be difficult to obtain. In relation to simulation studies, previous works [8], present interesting results concerning the use of MP2P architectures by using epidemic dissemination of data, resulting in high ratio of successful delivery. By using the storage backup nodes, the potential is to lower the packet delivery failure ratio and data corruption.

## III. MOBILE BITTORRENT PROTOCOL

The Bittorrent protocol is a peer-to-peer file sharing protocol. The protocol is more efficient for the transfer of large amounts of data *(usually in the hundreds of megabytes)*, rather than smaller ones. It differs from other P2P techniques, as pieces of a file are divided between peers who enter a network and then exchanged in order to complete a file transfer. This allows peers with low bandwidth to participate in large data transfers.

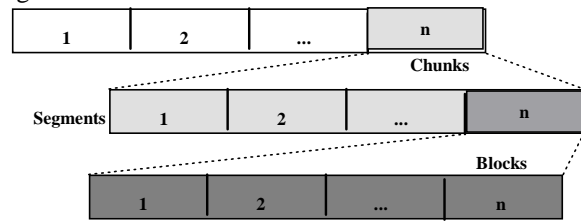

Figure 1: Blocks are the small pieces of data, made up of a few bytes, which are requested by peers. A serialization and reassembly of the blocks received constitutes a piece of the file.

Figure 1 shows the partitioning of a file, cut down compatibly in smaller sections, to be made ready for transfer from one peer to another. This process, known as swarming [1], allows peers in the same network to exchange these pieces *en masse*. The peers use different techniques to make sure they do not receive pieces they already have. If redundancy will occur, the network's latency will be dramatically increased, while throughput would drop.

The Bittorrent approach enables P2P systems to share efficiently any requested resources. However in a mobility-based framework, many different restrictions come to degrade the end-to-end availability. The clustered swarming technique allows mobile peers to exchange data more reliably than other P2P schemes [5], such as Kazaa, Gnutella and DC *(Direct Connect)* [16][17][18]. Though, due to the limited bandwidth capabilities which are encountered in wireless devices, the utilization of the Bittorrent protocol often becomes problematic.

*A. Current Problems in the existing static framework*

Problems of the protocol include increased latency while transferring small files, bandwidth problems, content unavailability, and leeching. One of the major inefficiencies of the protocol arises from a disproportionate distribution of content among peers, discussed in [9]. This kind of distribution allows peers to get different pieces from each other, which optimises the download/upload rates between seeders and peers, but it also holds the potential of breaking a swarm, since the piece holders may not exist in the network at all times. This is not as common with large swarms; however strategies are needed to promote smaller network sizes for improved delay and maximisation of throughput.

When a client first enters a swarm, they need to prove to their neighbours and other peers, as per-the-protocol's specification, that they are trustworthy enough in order to share information with. In order to achieve this, a small trial period of some minutes may pass, in which peers treat new peers with a bias, passing smaller amounts of data to them until they can prove that they will seed back what they've been given [1][3]. After this process takes place, peers start receiving much larger amounts of data. It is understandable therefore, that for smaller files it would not be worth sharing them through Bittorrent.

The transferring of data via the BitTorrent protocol puts a heavy load on the peers' bandwidth, observed by [10]. Since peers use a metadata file to locate pieces they need to download, the actual exchange of data is peer-to-peer and therefore a server is not involved. Thus, the bandwidth load occurs always on the client side and this is the main reason that service providers are opposing the use of the protocol.

There are often cases where a file is not as popular as others. When this is the case for extended periods of time, peers may not see the need to continue sharing this specific file, and therefore the seeding swarm dies out as per the lifetime scenarios of [3]. Content unavailability is a concept which is difficult to find easy solutions to. Even if an archive is unpopular, its value is many times unquestionable and therefore the archive needs to remain in circulation, especially if it is of scientific importance as many foundations may use P2P protocols such as Bittorrent to share these types of data. The essence behind this lays in the fact that even though peers still have the files stored on their storage media, they stop having them available to share in order to save bandwidth. However, even though valuable bandwidth is saved, peers entering a swarm to share an unavailable file will never be able to complete the transfer and the swarm will remain incomplete indefinitely or eventually die out.

One of the purposes of our algorithm aims in eliminating the above phenomenon through the implementation of partial mobility characteristics. This will allow wireless devices to evolve in a swarm through a higher download ratio even though their uploading bandwidths are not as high.

*B. Hybridised Model with Partial Mobility Characteristics using the Mobile Bittorrent Protocol*

The proposed model takes into consideration the difficulties which mobile clients, e.g., wireless devices, face while transferring files from other peers, most often static ones, through a Bittorrent network. Whilst the protocol offers an efficient way to share and distribute content, it has heavy requirements on bandwidth towards the client side. Content distributors benefit from peers using the protocol as they do not need to spend on acquiring large bandwidths and servers to distribute their content; rather only peers spend their bandwidth and CPU power to distribute the content. This is one of the reasons which internet service providers are often congested due to Bittorrent traffic. Users may not realize this, as the protocol makes it rather easy to share; however when it comes down to several network metrics, it is easily observable that content distributors benefit more than clients. In order to lay the grounds for a more efficient experience for the users, many of the problems described should be faced by devising the appropriate functionalities while not violating or altering the Bittorrent protocol.

Our algorithm also presents a way to control the latency between mobile and static peers. If mobile clients request data from other peers, the peers have the option of opening more connections, therefore serving more mobile peers at once. The reason of performing this, is because mobile peers have smaller bandwidths and limited connectivity, hence another peer may split their uploading activities between other mobile peers into greater than the default amount of connections allowed. A static peer is a non-moving peer or a normal peer itself; however it has larger bandwidth capabilities and therefore can provide more simultaneous connections, given that it transfers to mobile peers. This helps decongest not only the arrival requests from mobile clients, but also the network itself. As previously discussed, a Bittorrent seeder may upload to five connections at the same time. By implementing our strategy, peers with high latency issues will drop connections with specific peers in order to allow the faster seeders continue the transfer.

In the proposed model both static and mobile peers communicate with the tracker on a similar level; however, the tracker makes different kinds of decisions based on what type of peers the requests are coming from. For instance, the tracker decides how many uploading connections a seeder may open, by manipulating metainfo about the downloader's bandwidth limitations, instead of maintaining a default number of connections that it can open. The tit-for-tat policy is still implemented as our model does not violate any of the BitTorrent protocol aspects.

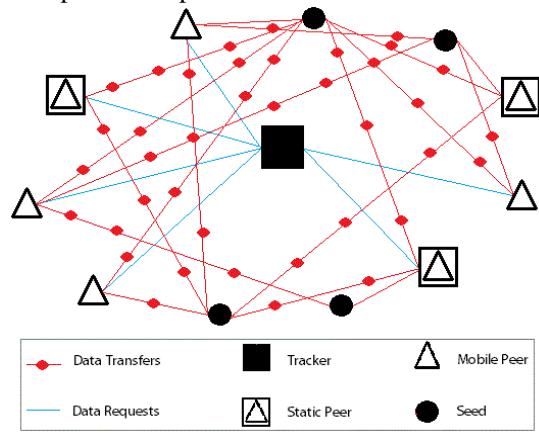

Figure 2: A showcase of the proposed model, presenting the distinction between mobile and static peers.

When a mobile client makes a request for bits, the tracker acknowledges the request by mapping more connections with available seeders. If the downloading section of the mobile peer's bandwidth is congested, the peer will ask the tracker to map fewer connections towards it. By using this technique, a client can ensure that their connection limitations are being used appropriately. In figure 2 it can be seen that seeders are allowed to share towards more mobile peers than static peers, whilst their uploading bandwidth is split equally between static and mobile peers. Bandwidth limitations still apply; therefore non-mobile clients with low bandwidth may not participate in seeding towards mobile peers. On the other hand, to ensure P2P fairness, mobile peers with a high bandwidth ratio may not be regarded as mobile per se.

Our model may be summarized in an algorithmic fashion for better understanding of the implementation, as shown in Figure 3.

```
Get swarmSize(N Peers);
Get_announce(peer, peer_type);
//tracker keeps track of peer_type in tables
Set_peer(peer, index);
//tracker indexes peer
Find_unchoked_peer();
Block_request(sourceIndex, destIndex)
   if destIndex.mobile == true
   {
      ConnectionSize.mobile = sourceIndex.uploadSize / destIndex.Size;
      while (count != ConnectionSize)
      {
         new Connection(sourceIndex, destIndex, block);
         count++;  //increase counter to check if max       connection size based
on bandwidth restrictions has been reached
      }
   }
   else
   {
      ConnectionSize = default; //default = 5.
      while (count != ConnectionSize)
      {
         new Connection(sourceIndex, destIndex, block);
         count++;  //increase counter to check if max allowed connection size has been reached
      }
   }
Send_blocks(Connection);
```

Figure 3: Pseudocode of the proposed hybridisation model with partial mobility characteristics.

## IV. SIMULATION RESULTS AND PERFORMANCE ANALYSIS

The proposed model's algorithm works without altering the BitTorrent protocol itself, but rather by implementing it on the client side and taking into consideration the dynamic changes in topologies. The use of seeders are proposed, who will have the ability to split their bandwidth capabilities and seed more mobile nodes at once, i.e., when there is a high latency from peer to peer transfer or when mobile peers are not getting a fair share of the content. This way, the overall latency of the transfers will be dropped significantly, thus lifting the bandwidth burden off the clients and allowing the content distributors give benefit to their users, both mobile and not.

A simulation was set up, running both a Bittorrent swarm with seeders who serve normally and a swarm in which seeders could serve more mobile peers simultaneously. The implementation-simulation of the proposed scenario was performed in Java programming language libraries as in [8]. We assume a system consisting of several mobile nodes, e.g., mobile users equipped with notebooks or PDAs and wireless network interfaces and that all devices are following a human-based activity (movements of nodes according to real-time pathways (roads, streets, corridors etc)). Radio coverage is small compared to the area covered by all nodes, so that most nodes cannot contact each other directly. Additionally, we assume IEEE 802.11x as the underlying radio technology.

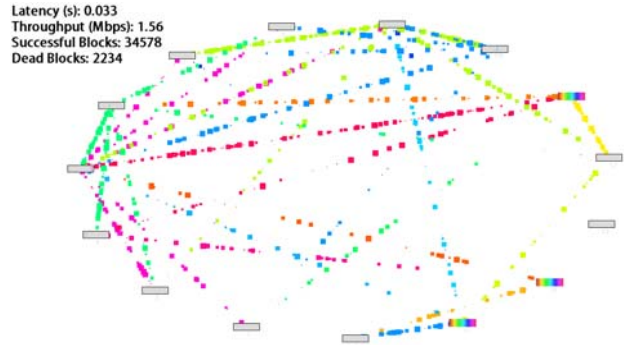

Figure 4: The simulation of the proposed model as viewed with graphical modes in order to enable visual representation of the Bittorrent resource sharing connectivity.

The simulation, presented in a visual format as seen in Figure 4, to further enable us to understand the techniques in which peers use the protocol to share data, derived metrics such as *average latency,* and *throughput*, since these are the targets of our model. Also, in conjunction to these, the *running time* for completing swarms of ratio 10 peers to 1 seeder was measured. In the simulation of the swarm with mobility characteristics, one seed was serving both mobile and non-mobile nodes, while a second was serving only non-mobile peers.

### A. Experimental Results

As figure 5 depicts, by comparing both the latency and runs of a normal Bittorrent swarm and a swarm which enables partial mobility characteristics, there is a difference in the latency response. In both cases the swarm contained the same number of peers, and the same number of seeders.

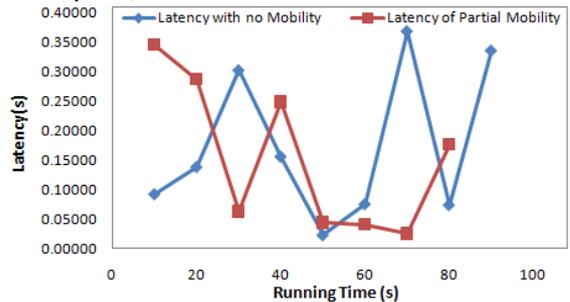

Figure 5: Comparing latency without partial mobility and latency with the running simulation time.

In the proposed algorithm, it is easily recognizable that the average latency from block-to-block has been decreased, due to the seeder being able to fill in the blanks, allowing more simultaneous mobility.

When referring to latency, we speak of block to block latency and not the initial lag which a peer experiences when connecting to a swarm. The reason for this is to minimize the transfer delays and therefore help the overall running time of the download. Equation 1 evaluates the delays from block to block.

$$T_\delta = K(t_x) - K(t_0) \quad (1)$$

where $T_\delta$ is the change in time from block to block, $K(t_x)$ the time a block has been released, and $K(t_0)$ the time a new block starts travelling towards destination.

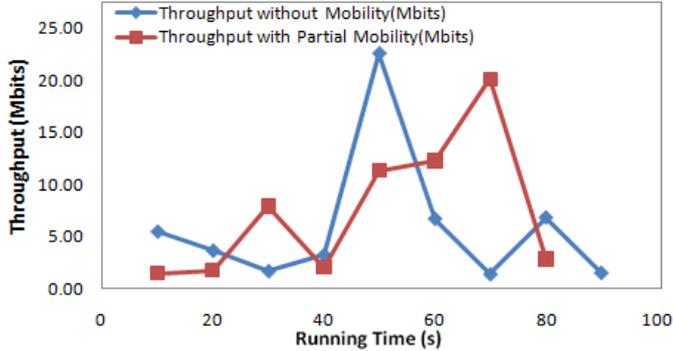

Figure 6: Throughput versus time for both partial mobility characteristics and no partial mobility characteristics in swarm clusters.

Whilst the throughput from peer to peer seems to peak in a normal swarm, this is only momentarily. As observed by the results extracted in Figure 6, the throughput of a swarm with partial mobility has a higher average through time, especially since partial mobility allows for a shorter running time in a network with both mobile and static peers. Formula in equation 2 shows how the throughput is represented.

$$C_{avg} = S / T_\delta \quad (2)$$

where $C_{avg}$ is the average throughput, and $S$ represents the number of blocks which have successfully reached their destination at any given point in time.

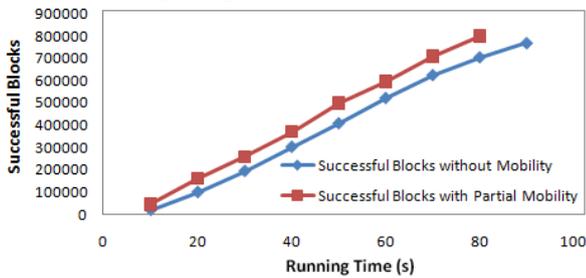

Figure 7: The number of gradual successful blocks through time for each simulation run.

At 50$s$, a swarm which has the ability of allowing seeders to upload to multiple mobile clients, may deliver 10% more blocks than a swarm without mobile characteristics.

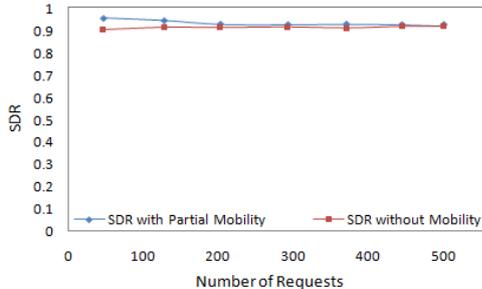

Figure 8: SDR with the number of requests occurring in Bittorrent resource sharing connectivity.

While our model is not expected to behave satisfactorily on small scale swarms, it seems to be effective for large scale transfers, minimizing the network's overall latency while increasing throughput from peer to peer.

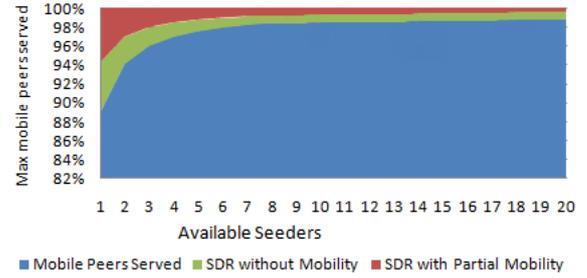

Figure 9: Outlining the maximum percentage of mobile peers which can be served based on the number of available seeders in a swarm.

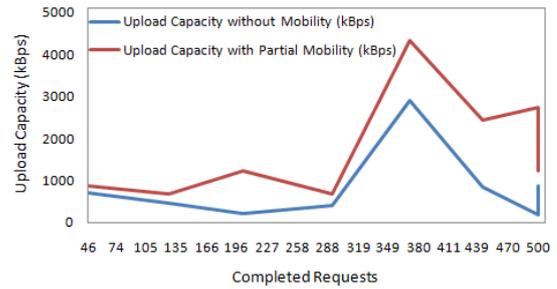

Figure 10: The upload capacity in kilobytes is given as a function of the completed requests. The upload capacity of a network with partial mobility characteristics is significantly higher per request than one with no mobility.

Additionally, not just content distributors would use this method, but any client could make use of the model. Of course, as in the BitTorrent protocol, some limitations still apply. The peers are still expected to keep uploading once they have acquired the entire file in order for the proposed method to work. Nevertheless, this will make it easier for peers to stay connected, since their bandwidth will only be used when other peers experience traffic problems.

### B. Optimisation Techniques

#### 1. Peer Selection

Even though some of the existing policies, such as *random piece first* and *rarest piece first*, are working on sufficient levels, the selection strategies for a swarm containing mobile peers cannot be maintained by simply these two techniques. The *choking* algorithm is a peer selection strategy which prefers clients with the highest upload rates [11], and this could work greatly towards the advantage of mobile peers in a swarm which uses our model, as naturally the non-mobile nodes hold the highest upload rates.

When peers finish downloading a file, thus becoming seeders, considering first that they have the available bandwidth, they can open more connections than the default, which is 5; however their uploading bandwidths will be split over those connections. This is done to allow mobile peers to enjoy downloading resources. Seeders could take turns in seeding towards mobile peers. On a similar note, if a peer has the uploading bandwidth to serve multiple nodes fast and efficiently, then they would be beneficial to the swarm, minimizing the latencies from block to block transfers. This follows well with the *observed upload rate* (OUR) in [12], which gives priority to peers which can upload data to other peers in a fast and reliable way.

Unlike the *LiveSwarm* protocol found in [13], our model does not need the seeder to push data to other peers. As we have observed from our experiments, Bittorrent's standard method of peers requesting data works more robustly than pushing for a few reasons. The first reason is that clients who are already choked or even who want to appear choked are not given the possibility of doing so. Secondly, as discussed in previous sections, the proposed model attempts to prevent seeders from uploading faster than mobile peers can download. Thirdly, the Bittorrent protocol does not need to be modified in order for our algorithm to work, as our model changes only the information the tracker and peers exchange.

*2. Seeding Strategies*

As in [14], optimistic unchokes would not be needed if nodes were able to calculate the upload bandwidth for the peers servicing it. In our model, since all peers communicate continuously with the tracker which constantly updates the metainfo it receives, a node could receive such bandwidth statistics from the tracker, thus eliminating the need for optimistic unchoking between peer and static seeder, and performing the unchoking algorithm only for non-static peers.

*3. Efficient Distribution of Data*

Concerning the notion of distributing data efficiently, suggestions show the importance of delay-sensitive responses to peer requests [15]. Through the use of such defensive measures taken by seeding peers, the broadcasting of data may be efficiently redistributed by nodes which can make use of multicast technologies. Decisions should be made based on querying the neighbouring peers in a Bittorrent swarm, and through the collection of these feedbacks in order to create the appropriate responses.

## V. CONCLUSION AND FUTURE WORK

In this work, a new model concerning the involvement of P2P strategies with partial mobility characteristics was proposed, where clients in a network adopt techniques to seed more efficiently to mobile nodes. The round trip delays were considered and strategies for peer selection and seeding policies were suggested. We have entailed the potential of partial mobility characteristics in a peer-tracker and peer-peer communication schema, which allows an optimized approach in attaining high resource availability and lower packet failure ratios in mobile transfers through the use of Bittorrent.

In order to assess our own model, we needed to take some measurements of the Bittorrent protocol on the specific metrics our algorithm proposes to improve. This was through the comparisons of the experimental results of our model and the existing protocol. Our results show a satisfactory improvement in delay times and the increase of throughput. Likewise, we have observed that our model provides faster runs between mobile and static peers.

Future research directions include the implementation of the seeding strategies and peer selection techniques. Moreover, the combination of other mobility schemas with our model gives the potential to create a truly mobile Bittorrent implementation.